# Unexpectedly super strong paramagnetism of aromatic peptides due to cations

Haijun Yang[1,2], Liuhua Mu[2], Lei Zhang[3], Zixin Wang[4], Shiqi Sheng[5], Peng Xiu[6], Yongshun Song[5], Jun Hu[1,2], Xin Zhang[3*], Feng Zhang[4,7*], Haiping Fang[5,1*]

1. Shanghai Synchrotron Radiation Facility, Zhangjiang Laboratory (SSRF, ZJLab), Shanghai Advanced Research Institute, Chinese Academy of Sciences, Shanghai 201204, China
2. Shanghai Institute of Applied Physics, Chinese Academy of Sciences, Shanghai 201800, China
3. High Magnetic Field Laboratory, Key Laboratory of High Magnetic Field and Ion Beam Physical Biology, Hefei Institutes of Physical Science, Chinese Academy of Sciences, Hefei, China
4. State Key Laboratory of Respiratory Disease, Guangzhou Institute of Oral Disease, Stomatology Hospital, Department of Biomedical Engineering, School of Basic Medical Sciences, Guangzhou Medical University, Guangzhou 511436, China
5. School of Science, East China University of Science and Technology, Shanghai 200237, China
6. Department of Engineering Mechanics, Zhejiang University, Hangzhou 310027, China
7. Biomedical Nanocenter, School of Life Science, Inner Mongolia Agricultural University, 29 East Erdos Street, Hohhot, 010011, China

*Corresponding author. Email: fanghaiping@sinap.ac.cn; fengzhang1978@hotmail.com; xinzhang@hmfl.ac.cn

Abstract:
**We found that the AYFFF self-assemblies in the chloride solution of some divalent cations ($Zn^{2+}$, $Mg^{2+}$, and $Cu^{2+}$) display super strong paramagnetism, which may approach the mass susceptibility of ferromagnetism. We attribute the observed super strong paramagnetism to the existence of the aromatic rings, which interact with the cations through cation-π interaction.**

It has been known for centuries that a large passel of animals navigates using the Earth's magnetic field. Later, scientists found that cell orientation, proliferation, microtubule and mitotic spindle orientation, DNA synthesis and cell cycle can be significantly affected by high static magnetic fields [1-4]. For example, Tendler and Lee found that self-assembled aromatic peptide nanotubes could align in high magnetic fields, [5, 6] which indicates that the alignment of aromatic peptide nanotube in magnetic fields was mainly originated from the ordered structure of aromatic rings in the peptide nanotube, given that the aromatic rings have a large diamagnetic anisotropy.[5, 6]

Although the effects of high magnetic field on some biological samples have been clearly demonstrated, it is still unclear that how living organisms reliably sense weaker magnetic field, such as geomagnetic field (~50 μT), in the presence of thermal fluctuations and other sources of noise. In general, most biological systems have very weak diamagnetic or paramagnetic properties. Whether the magnetoreception in animals originates from magnetic particles is still under considerable debate. In 2015,

Qin et al. reported a new putative magnetic receptor MagR that is potentially ferromagnetic [7]; however, this finding subsequently was challenged by Meister, who argued that the energetic interaction of the geomagnetic field with such magnetoreceptor is too small (by at least five orders of magnitude) to overcome thermal fluctuations at room temperature [8]. Moreover, it was proposed that magnetite in the upper-beak skin of homing pigeons is responsible for magnetoreception, but the subsequent works argued that it was macrophage contaminations[9, 10]. Meanwhile, there have been efforts trying to understand the biomagnetism of organic components without ferromagnetic components, such as the radical-pair mechanism with cryptochromes. However, there is still no fully satisfactory answer[11, 12], as mentioned in the very recent paper, "Does Quantum biology exist (that is, is magnetic sensing truly quantum in at least some animals)?"[13]

Moreover, the presence of water in living organisms have completed the situation, which was shown to affect the magnetism of biological samples. It has been demonstrated that non-dehydrated DNA show a paramagnetic upturn at low temperature, while dehydrated DNA is completely diamagnetic. [14] In addition, we recently showed that calcium ions on graphene showed ferromagnetic properties[15, 16], in which the key is the cation-π interactions between the calcium ions and the polycyclic aromatic rings on graphene, and the calcium ions show monovalent behavior in CaCl crystals on graphene, and this mechanism is also applicable for other divalent cations. Therefore, we hypothesize that if aromatic rings are enriched in a biomolecule, for example, biopolymers, the divalent cations absorbed on the polymers may also display strong paramagnetic and even ferromagnetic properties.

In this paper, we take the AYFFF aromatic peptides as an example, while using IIIGK as negative control (Figure 1a). We found that the AYFFF self-assemblies in the $ZnCl_2$, $MgCl_2$ and $CuCl_2$ solutions of sufficiently high concentrations display super strong paramagnetism, which may even approach the mass susceptibility of ferromagnetism. We attribute the super strong paramagnetism to the existence of the aromatic rings, which interact with the cations through cation-π interaction.

In our experiment, the AYFFF aromatic peptide powders were first dispersed into pure water (Milli-Q, 18.2 MΩ) with the concentration of 0.5 mg/mL, and stored still at 20 ℃ for 3 days. Some dispersions were thoroughly mixed with $ZnCl_2$ solutions to obtain mixtures with different $ZnCl_2$ concentrations. After settling for 30 minutes, the supernatant was taken out for the characterization of morphology and magnetic susceptibility. Figure 1d displays the typical self-assembled AYFFF peptide microfibers in the supernatant, which are longer than 2.0 μm with the height of ~1.5 nm. For comparison, some dispersions were thoroughly mixed with pure water, in which no cations exist.

We used a Quantum Design MPMS3 SQUID magnetometer to measure the direct current (DC) magnetic susceptibility at 298 K. Peptide mixtures of about 160 μl were loaded into a liquid sample holder (C130D, Quantum Design) and sealed tightly to avoid leakage in vacuum. The magnetic field was swept between -30,000 Oe and 30,000 Oe with one measurement point per 2500 Oe. All DC magnetic susceptibilities were corrected for diamagnetic contribution from the sample holder, and the solvent (water or salt aqueous solution) by directly subtracting their scaled voltage signals by weight and then fitting with a SquidLab program [17]. As the result, we can get the magnetization (*M*) versus magnetic field (*H*) curves, and compute the mass susceptibility ($\chi$), $\chi = M/H = a/mH$, where *a* is the moment measured by the SQUID magnetometer and *m* is the mass of the assembled peptide measured by a Nanophotometer$^{TM}$ P330 (Implen GmbH).

Figure 1b displays a typical example of the magnetization (*M*) measured with respect to the magnetic field (*H*) applied for the AYFFF peptide assemblies in 40 mM $ZnCl_2$ solution (purple dots). We linearly fit the magnetization (*M*) with respect to the magnetic field (*H*). The slop is $2.32 \times 10^{-4}$ emu/g, which is three orders of the magnitude larger than the absolute value of the susceptibility of water, reaching the susceptibility of superparamagnetism. In Fig. 1c, we show the mass susceptibility ($\chi$) for the peptide for four independently repeated trials. The $\chi$ ranges from $1.68 \times 10^{-4}$ emu/g to $2.98 \times 10^{-4}$ emu/g, with the average value of $2.47 \times 10^{-4}$ emu/g in the $ZnCl_2$ solution of 40 mM.

We also measure mass susceptibility ($\chi$) for the peptide by weighing the magnetic force of peptides at different state under a fixed magnetic field by using a high-precision electronic balance (XPR205, METTLER TOLEDO, Switzerland) as described in Ref .[18]. [18] The results are also shown in Fig. 1c (pink dots). The mass susceptibility ranges from $1.41 \times 10^{-5}$ emu/g to $2.97 \times 10^{-4}$ emu/g with the average value of $1.61 \times 10^{-4}$ emu/g, which is close to the data obtained by the SQUID magnetometer. We have further measured the mass susceptibility for the peptide in the other chloride solutions of $MgCl_2$ and $CuCl_2$ with the same concentration of 40 mM. As shown in Fig. 1c, the mass susceptibility of the peptide in salt solutions of $ZnCl_2$ and $MgCl_2$ are comparable and the mass susceptibility for $CuCl_2$ is considerable larger, suggesting that the super strong paramagnetism of the assembled nanofibers of AYFFF in the chloride solution of the divalent metals should be universal.

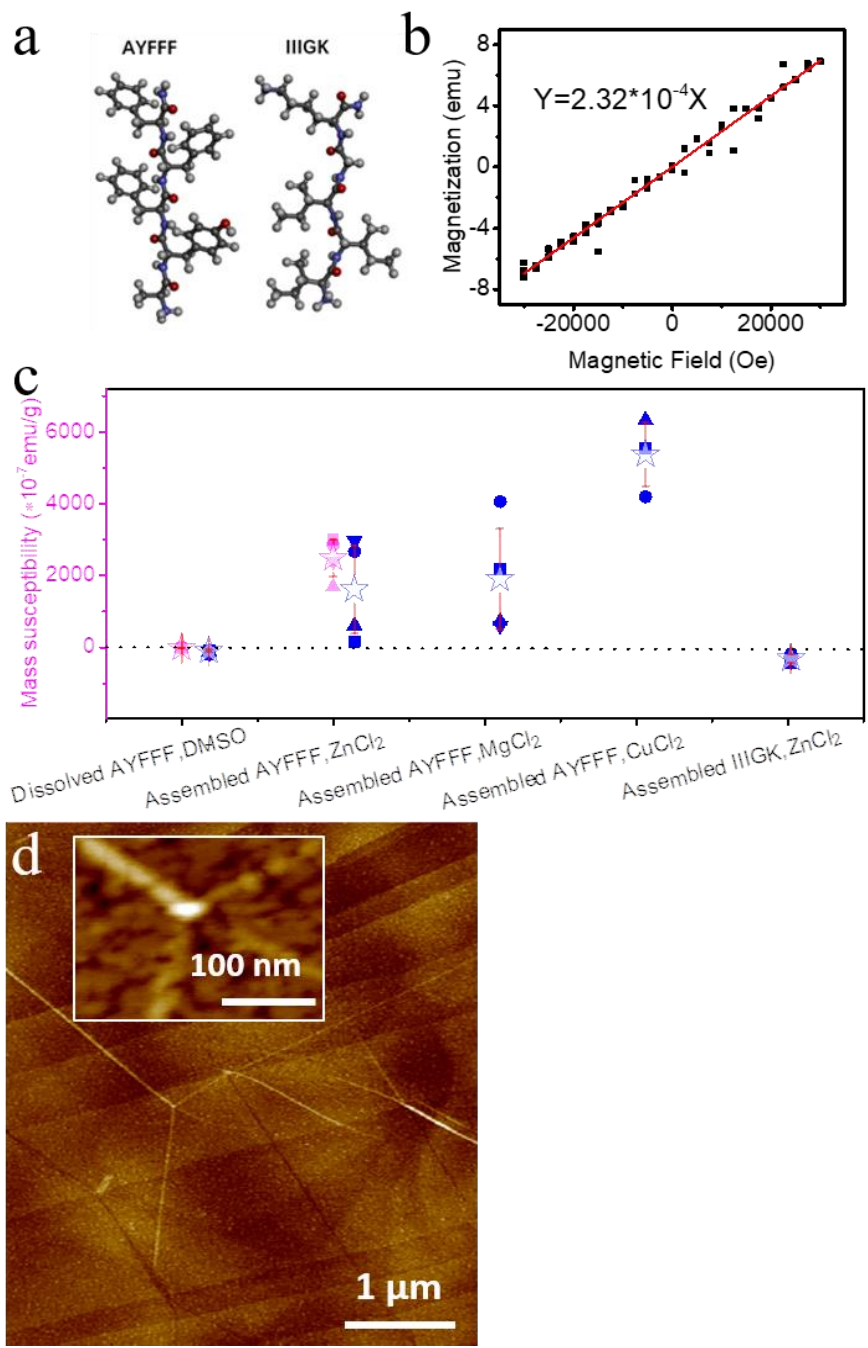

Figure 1. *a) Molecular structures of the AYFFF aromatic peptide and the IIIGK nonaromatic peptide. Note: both peptides are synthesized with the C-terminal amidation in order to increase their biological activities via generating a closer mimic of native proteins. b) Magnetization (M) versus magnetic field (H) of AYFFF peptide assemblies dispersed in the $ZnCl_2$ solution of 40 mM. c) Average mass susceptibility of AYFFF peptide at dissolved, and assembled states in different solutions, together with non-aromatic IIIGK as a control. The salt concentrations of $ZnCl_2$, $MgCl_2$, and $CuCl_2$ are 40 mM. The purple and blue solid scatters display the experimental data collected by the SQUID magnetometer and the weighing method described in Ref.[18], respectively. The hollow stars represent their average values with error bars showing*

*their standard deviations. d) AFM images of the self-assembled microfibers of AYFFF aromatic peptides form the supernatant of the mixture with the ZnCl$_2$ solution of 40 mM.*

As mentioned in the introduction, we presumed that the aromatic rings are the origin of the super strong paramagnetism. As a negative control, we also performed measurements with a nonaromatic peptide with the sequence of IIIGK. With the same methods as used for AYFFF peptides, we obtained the mass susceptibility of IIIGK in the ZnCl$_2$ solution of 40 mM, with an average value of $-3.2 \times 10^{-5}$ emu/g. This confirms that the presence of aromatic rings is the key reason that contributed to the superparamagnetic properties of AYFFF, and suggests that ions endowed AYFFF assemblies with super strong paramagnetism through the similar mechanism of cation-π interaction reported in Ref. [15].

In summary, the AYFFF self-assemblies in the ZnCl$_2$, MgCl$_2$ and CuCl$_2$ solution of sufficiently high concentrations display super strong paramagnetism, which may possibly approach the mass susceptibility of ferromagnetism. We attribute the super strong paramagnetism to the existence of the aromatic rings, which interact with the cations through cation-π interaction with the mechanism reported in Ref. [15]. Therefore, we expected other peptides, proteins or other compounds that contain higher percentages of aromatic rings in the solution of other divalent metals with sufficiently high concentrations could also have high paramagnetism. Our findings provide an important step towards the understanding of not only the magnetism of biomolecules and the origin of magnetoreception in living organisms, but also the magnetic effects and magnetic controls on aromatic ring-enriched biomolecule and drugs, including their dynamics and reactions, in living organism.

**Acknowledgement**
This work is supported by the National Natural Science Foundation of China (Nos. U1632135, 11974366, U1932123, 51763019, U1832125 and 31900506), the Key Research Program of Chinese Academy of Sciences (Grant No. QYZDJ-SSW-SLH053), the Fundamental Research Funds for the Central Universities, the National Science Foundation of Anhui province (1908085MA11), and the Anhui Postdoctoral Foundation (2019B367).